\documentclass[dvips]{article}

\usepackage{icrc2011}

\title{MAGIC measurement of the Crab Nebula spectrum over three decades in energy}
\newcommand{\etal}{\MakeLowercase{\textit{et al. }}} % "et al."
\shorttitle{Zanin, R. \etal Crab Nebula spectrum obtained with MAGIC stereo.}

\authors{Zanin, R.$^{1}$, Mazin, D.$^{1}$, Carmona, E.$^{2}$, Colin,
  P.$^{3}$, Cortina, J.$^{1}$, Jogler, T.$^{3}$ Klepser, S.$^{1}$, Moralejo, A.$^{1}$, Sitarek J.$^{4}$, for the MAGIC Collaboration, 
and Horns, D.$^{5}$ and Meyer M.$^{5}$}
\afiliations{$^1$Intitut de Fisica d'Altes Energies, Bellaterra, Spain\\
$^2$Centro de Investigaciones Energ\'eticas, Medioambientales y Tecnologicas, Madrid, Spain\\
$^3$Max-Planck Institut, Munich, Germany\\
$^4$University of \L\'od\'z, Poland\\
$^{5}$University of Hamburg, Germany}
\email{roberta@ifae.es}

\abstract{
The Crab Pulsar Wind Nebula is the best studied
source of $\gamma$-ray astrophysics. The contribution of the various
soft radiation fields to the Inverse Compton component of its high
energy emission, the strenght of the internal magnetic field and the
maximum energies reached by primary electrons are however still matter
of study.\\
The MAGIC stereoscopic system recorded almost 50 hours of Crab Nebula
data in the last two years, between October 2009 and April
2011. Analysis of this data sample using the latest improvements in
the MAGIC stereo software provided an unprecedented differential
energy spectrum spanning three decades in energy, from 50 GeV up
to 45 TeV.
At low energies, the MAGIC results, combined with the Fermi/LAT data,
yield a precise measurement of the Inverse Compton peak. 
In addition, we present light curves of the Crab Nebula at
different time scales, including a measurement simultaneous to one of the
Crab Nebula flares recently detected by both Fermi/LAT and AGILE. Using the
MAGIC spectrum together with multiwavelength data, we discuss the
implications for the modeling of the Crab Nebula. 
}


\begin{document}
\maketitle

%Begin the section.
\section{Introduction}
The pulsar wind nebula (PWN) associated with the Crab Pulsar is a 
leftover of a supernova explosion recorded in 1054
\cite{Stephenson2002}. Located at the center of the nebula, the pulsar
continuously releases its energy primarily in the form of a highly
relativistic and magnetized wind of particles, mainly electrons and
positrons. This pulsar wind terminates in a standing shock where
particles are accelerated up to ultra-relativistic energies, and their
pitch angles are randomized. The outflow interacts with the
surrounding magnetic and photon fields creating the PWN. The nebula
emits synchrotron radiation which is observed from radio frequencies
up to soft $\gamma$-rays. This emission is well-described in terms of
the magnetohydrodynamic model (MHD) \cite{KennelMHD}. At higher
energies (above 500 MeV), the overall emission is dominated by the Inverse
Compton (IC) scattering by the electrons from the pulsar, predominantly on 
the synchrotron photons \cite{Atoyan1996}. 

The Crab Nebula is one of the best-studied non-stellar objects in the sky.
Due to its bright glow at almost all wavelengths, precise measurements
can be provided by many different kinds of instruments, allowing for a
detailed examination of its physics.\\
The IC emission from the Crab Nebula was detected for the first time
above 700 GeV by the pioneering Whipple telescope in 1989
\cite{Weeks1989}. Since then, the imaging Cherenkov technique was
successfully used to extend the Crab Nebula differential energy
spectrum from a hundred GeV up to a hundred TeV. Nevertheless, the
low-energy part of this component, below 100 GeV, where the IC peak
resides, could be observed only in the last years. At low energies,
the \emph{Fermi} satellite filled the gap between few and hundred GeV
\cite{Abdo2010}. In the meanwhile, the construction of imaging
atmospheric Cherenkov telescopes (IACTs) with larger
reflective surface allowed to lower the energy threshold of the
ground-based telescopes below 100 GeV. The stand-alone first MAGIC
telescope had already shown a hardening of the spectrum below a few
hundreds GeV \cite{Nepomuk}. With an energy threshold of 80 GeV, MAGIC
provided the only experimental data points overlapping with the
\emph{Fermi}/LAT ones up to now. The goal of this work is to measure
the Crab Nebula differential energy spectrum with a higher statistical
precision and down to lowest possible energy threshold by using the
MAGIC stereoscopic system.  

In addition, because of its apparent overall flux steadiness, the Crab
Nebula has been considered as standard candle at almost all
wavelengths. It has been used to cross-calibrate X-ray and
$\gamma$-ray telescopes, to check the instrument performance over 
time, and to provide units for the emission of other astrophysical
objects. However, this notion surprisingly changed in September 2010,
when both \emph{AGILE} and \emph{Fermi}/LAT satellites observed an
enhancement of the $\gamma$-ray flux above 100~MeV by a factor 2
\cite{Tavani2011, Abdo2011}. Spectral analysis revealed that the flare
had a synchrotron origin, and a spectrum which is harder than the
average one.  Optical and X-ray images taken right after the flare
showed a few nebular, brightened features. One of them is the
``anvil'', the knot which lies in the inner nebula in the projected
inward extension of the jet, and it is considered a primary site for
particle acceleration. If the $\gamma$-ray flare luminosity suggests
that the production region is close to the pulsar, the short flare
rising time favors a compact emission region of size smaller than
10$^{16}$ cm. Thus, the ``anvil'' feature could be an excellent flare
site candidate \cite{Tavani2011}.  
A similar episode was observed in April 2011 by the same teams
(ATel $\#$3276). It lasted for two weeks, with the peak occurring on
April 14, 2011. \\
The MAGIC telescopes recorded some data simultaneously to both
high-energy $\gamma$-ray flares in September 2010 and April 2011. The
results of these observations will also be presented in this paper.

\section{The MAGIC telescopes}
MAGIC consists of two 17~m diameter IACTs located in the Canary island
of La Palma, Spain (2200~m above sea level). It became a stereoscopic
system in Autumn 2009. The stereoscopic observation mode led to a
significant improvement in the performance of the instrument
\cite{Carmona2011}\cite{julian}. The current sensitivity of the array yields
5$\sigma$ significance detections above 250 GeV of fluxes as low as
0.8$\%$ of the Crab Nebula flux in 50 hr. The energy threshold of the
analysis lowered down to 50 GeV for low zenith angle observations. 

For this work we considered stereoscopic observations of the Crab
Nebula carried out during the last two years, between October
2009 and March 2011. These observations are performed in wobble mode,% \cite{Fomin1994},
at zenith angles between 5$^\circ$ and 50$^\circ$. Data affected by
hardware problems, bad atmospheric conditions,
or displaying unusual background rates were rejected in order to
ensure a stable performance. 48.7 hours of effective time are left
after data quality selection. 

The analysis was performed by using the tools of the MAGIC analysis
software \cite{Berger2011}. Each telescope records
only the events selected by the hardware stereo trigger. The so-called
image cleaning procedure selects the pixels which have significant
signal. The obtained reconstructed image is then parametrized with a
few simple quantities. In particular, we use a new algorithm, the
so-called \emph{sum image cleaning} to improve the analysis
sensitivity at energies below 100 GeV. Similarly to the standard
algorithm it uses the concept of core and boundary pixels, defined
according to their charge levels. The addition of new constraints on
the charge level of compact group of neighboring core pixels, within a
very short time spread, further allows to lower the charge thresholds.
This is an important aspect in the reconstruction of low-energy shower
images which contain only few pixels. The \emph{sum image cleaning}
performs as well as the standard one for energies above 150 GeV,
whereas below such energies it improves the sensitivity by a 15$\%$.

Afterwards, pairs of images from the same stereo events are
combined and the shower direction is determined as the average
of the corresponding single-telescope directions \cite{julian}.
The background rejection relies on the definition of the multi-variable
variable \emph{hadronness}, which is computed by means of a Random
Forest (RF) algorithm. RF uses as input a small set of image
parameters from both telescopes, together with the information about
the shower location provided by stereoscopy. The $\gamma$-ray signal is estimated through
the distribution of the squared angular distance ($\theta^2$) between
the reconstructed and the catalog source position. The energy of each
event is estimated by using look-up tables created from Monte Carlo
simulated $\gamma$-ray events. For the computation of the differential
energy spectrum, the $\gamma$-ray signals in each energy bin are
determined by selecting a soft \emph{hadronness} cut which retains 
90$\%$ of the $\gamma$-ray events.

\begin{figure}[!h]
\centering
\includegraphics[width=2.3in]{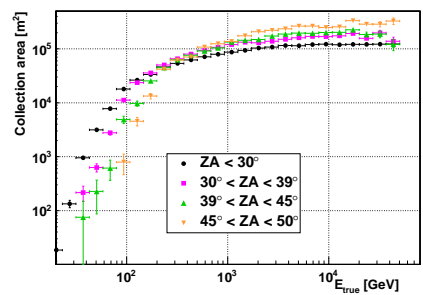}
\caption{Collection area obtained for the data sub-samples in the four different ranges of zenith angle.
\label{fig:Aeff}}
\end{figure}

%The use of the new image cleaning algorithm allows to reconstruct
%the first spectral point in the energy bin between 43 and 58 GeV with
%a Li $\&$ Ma (formula 17) significance of more than 3 $\sigma$. 
%Figure \ref{fig:SumIC} illustrates the distributions of the number of excess
%events and of significance per spectral point obtained with
%the standard and the \emph{sum image cleaning} analysis,
%respectively. The significance is computed by using the Li $\&$ Ma
%formula 17. 
%\begin{figure}[!h]
%\centering
%\includegraphics[width=2.5in]{icrc0195_fig01.eps}
%\caption{Distribution of the excess events and of the LiMa
%  significance per spectral point for the standard analysis and the
%  one with the sum image cleaning.
%\label{fig:SumIC}}
%\end{figure}
The data sample was divided in four zenith angle ranges to account for
corresponding variations in the image parameters. The matrices for the background
rejection obtained through the RF were computed for each sub-sample
separately. The four independent analyses were combined later at the
last stage of the analysis. In Figure \ref{fig:Aeff} we show the
collection area obtained for the four data sub-samples.

\section{The differential energy spectrum}
\begin{figure}[!t]
\centering
\includegraphics[width=3.3in]{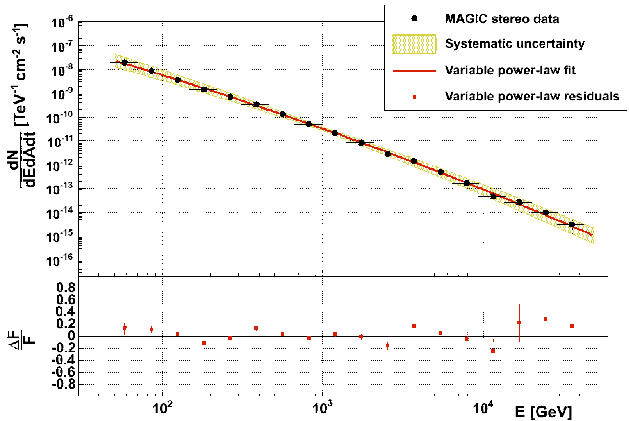}
\caption{Differential energy spectrum of the Crab Nebula obtained with
  data recorded by the MAGIC stereoscopic system. 
\label{fig:spectrum}}
\end{figure}

Figure \ref{fig:spectrum} shows an unprecedented differential energy
spectrum of the Crab Nebula which spans three orders of magnitude in
energy, from 50 GeV up to 45 TeV. In order to correct for the finite
energy resolution and energy estimation bias it was unfolded with
Bertero's method \cite{Bertero}. Three other unfolding methods were
considered, and all of them gave compatible results within the
statistical errors. 
The yellow-shadowed area indicates the systematic uncertainties on the
flux normalization and the spectral slope. The first one is estimated
to be 15$\%$ of the flux value, whereas the second one produces an
absolute uncertainty on the photon index of 0.15. 
%They account for
%different sources of systeatic effects: non-linearity of the analog
%signal chain, mis-pointing, discrepancies between data and MC,
%background rejection, zenith angle of the observations and different
%night-sky-background levels \cite{julian}. 
Additionally, there
is another systematic uncertainty affecting the energy scale
measurement at the level of 15--17$\%$. Systematic uncertainties,
related to the non-perfect matching of the observation conditions and
the Monte Carlo simulations, have been estimated by studying the
spectra reconstructed in different zenith angle ranges, analysis cuts,
mis-pointing values, night sky background levels \cite{julian}.
At all energies, the overall uncertainty is dominated by systematic
errors rather than statistical ones. 
We fit the differential spectrum with a variable power-law:
\begin{equation}
\mathrm{\frac{dN}{dEdtdA}} = \mathrm{f}_0
\left(\frac{E}{1\mathrm{TeV}}\right)^{- \alpha + b \cdot \mathrm{Log10} \left(
    \frac{E}{1\mathrm{TeV}} \right) } \mathrm{\frac{photons}{TeV cm^{2} s}}
\end{equation}
%\end{small}
where f$_0$ = $(3.27 \pm 0.03_{stat}) \times 10^{-11}$,
$\alpha$~=~2.40~$\pm$~0.01$_{stat}$, and
b~=~-0.15~$\pm$~0.01$_{stat}$. The residuals to the
fit show significant deviations which are explained by systematic
effects ($\chi^2/ndf = 72/15$).  

\begin{figure}[!h]
\centering
\includegraphics[width=3.3in,height=2in]{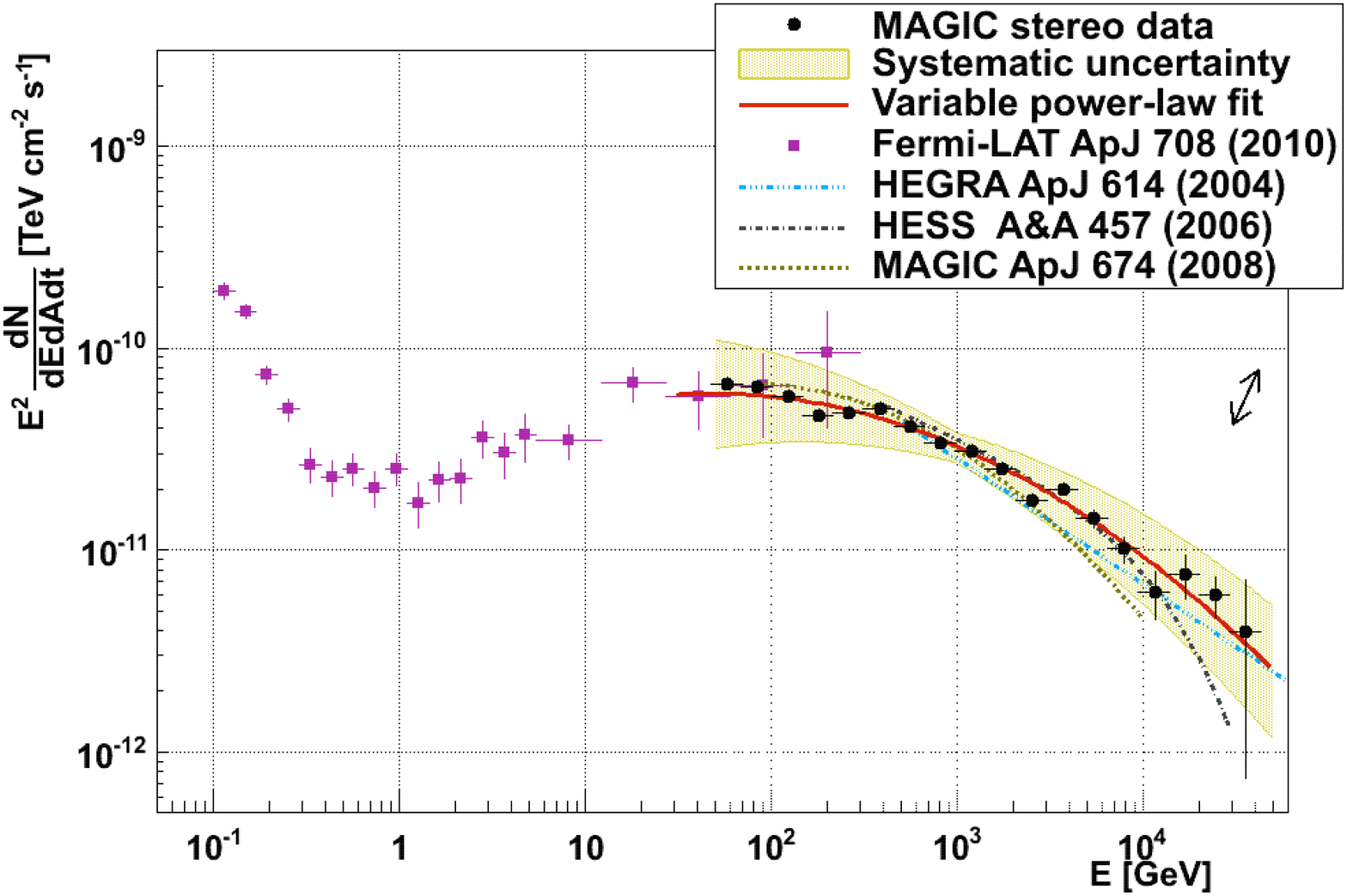}
\caption{Spectral energy distribution of the Crab Nebula obtained with
  the MAGIC telescopes, together with the results from previous
  $\gamma$-ray experiments. The black arrow indicates the systematic
  uncertainty on the energy scale. 
\label{fig:SED}}
\end{figure}

Figure \ref{fig:SED} shows the spectral energy distribution (SED) for
the same data points in Figure \ref{fig:spectrum}, and compares it to
other measurements by IACTs. At energies above
10 TeV, given the large systematic error, it is not possible to
disentangle the discrepancies between the results obtained by HESS
\cite{Aharonian2006} and those by HEGRA \cite{Aharonian2004}. 

In order to estimate the position of the IC peak we fit a
variable power-law function to our data combined with the
\emph{Fermi}/LAT data points taken from \cite{Abdo2010}. The fit
accounts for the correlation in MAGIC data point and considers
statistical errors only. The obtained IC peak is 59~$\pm$~6 GeV. 
\section{The light curve}
\label{sec:LC}
In this section we present a time-resolved measurement, i.e. the
light curve, of the $\gamma$-ray flux above 300 GeV from the Crab
Nebula. This is meant to check the flux stability on timescales of
days. %The extracted $\gamma$-ray signal is converted to flux by
%normalizing the $\gamma$-ray excess to the effective observation time
%and collection area in each time bin. 
 \begin{figure*}[th]
  \centering
  \includegraphics[width=6in]{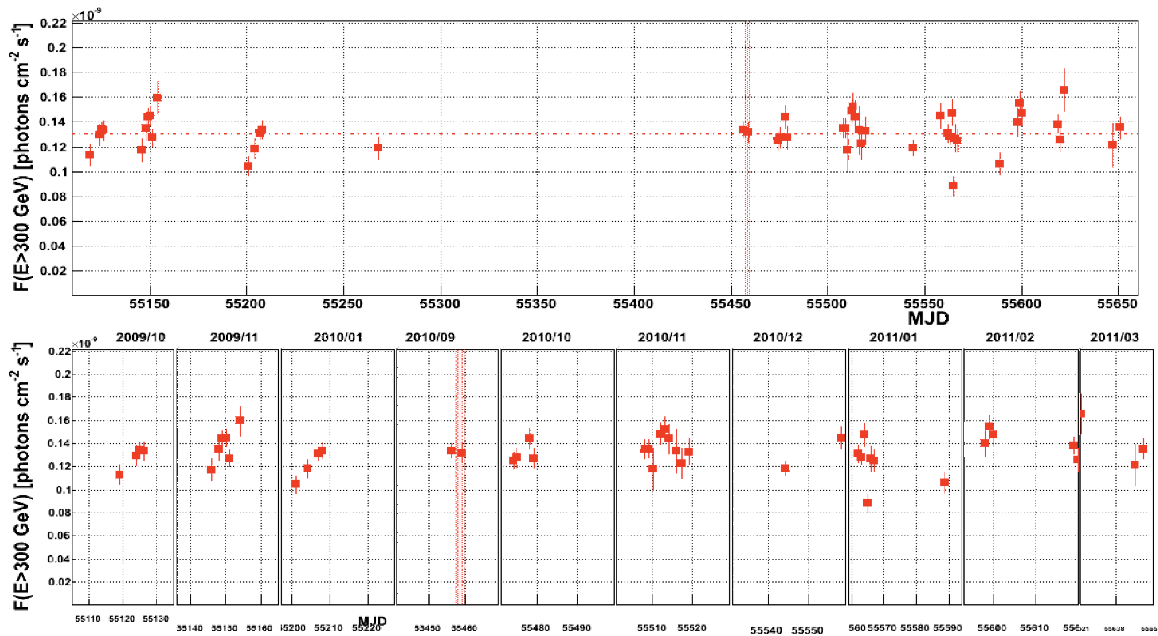}
  \caption{Light curve of the Crab Nebula.}
  \label{fig:LC}
 \end{figure*}
Results are shown in Figure \ref{fig:LC}. The upper
panel shows the light curve between October 1, 2009 and March 31,
2011. In addition, the bottom panels are zoomed per-month portions of the
bottom panel. The average flux of all nights
F$_{>300\mathrm{GeV}}$ is:
\begin{equation}
\mathrm{F}_{>300\mathrm{GeV}} = (1.31 \pm 0.03_{stat} \pm
0.17_{sys}) \times 10^{-10} \mathrm{cm^{-2} s^{-1}}.
\end{equation}
The Crab Nebula flux is stable within the systematic uncertainty
with a probability of 95$\%$.
The red shadow area in the panel marks the period of the enhanced
$\gamma$-ray flux detected by both \emph{AGILE} and \emph{Fermi} in
September 2010. MAGIC observed the Crab Nebula for
one night during the same period, on September 20, 2010. The obtained
flux is perfectly compatible with the average flux of all the nights
within the statistical error.  

\begin{figure}[h]
  \centering
  \includegraphics[width=3in]{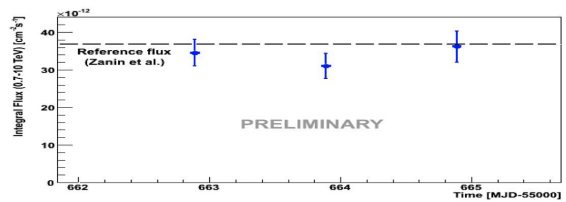}
  \caption{Light curve of the Crab Nebula between April, 12 and 14 
    during the flare detected at high-energies.}
  \label{fig:flare}
 \end{figure}

In addition to the sample used for the analysis presented up to here,
we took some data between April 12 and 14, 2011, for a total amount of
about 140 minutes. These observations were carried out under strong
moonlight conditions with a special high voltage setting. A
preliminary analysis of this second data set shows no increase in
flux, as illustrated in Figure~\ref{fig:flare}.

\section{Discussion}
\begin{figure}[h]
  \centering
  \includegraphics[width=3.3in,height=1.8in]{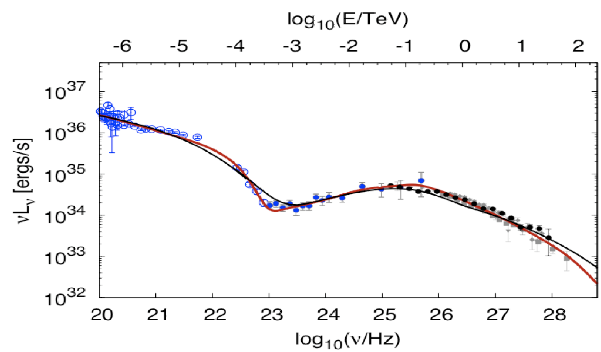}
  \caption{Measurements of the Crab Nebula SED (filled black circles
    are MAGIC data presented here and filled blue circles
    \emph{Fermi}/LAT points) overlaid to two
    different models: the constant B-field (red line) and the MHD
    (black line). 
  \label{fig:model}}
 \end{figure}

The broad-band SED of the Crab Nebula has been
matched with two different model calculations that are based on the
model first suggested by \cite{Hillas1998} assuming a constant magnetic field, B,
as well as a solution of the ideal radial MHD flow
\cite{KennelMHD}. The best match of the free model parameters (average
B-field and magnetization of the flow at the shock, $\sigma$) to the
measured SED has been obtained using a rather weak field of $B=(124
\pm 6) \mu G$ \cite{Meyer2010} or a value
$\sigma=0.0045\pm0.0003$. Both model calculations are shown in Figure
\ref{fig:model} together with the 
archival data as well as the new measurement presented here. The
spectral measurement obtained with the MAGIC telescopes covers for the
first time the crucial energy range at the peak of the IC
component. None of the models provide a satisfactory match to the
data if we consider statistical errors only. However, if taking
systematic uncertainties into account, both models can fit the
observations. In case the systematic errors of the
measurement can be further reduced, the new measurement can therefore
help to lift a degeneracy and will require the introduction of
additional structure in the B-field downstream of the
shock. Specifically, the shape of the spectrum at
$\approx 100$ GeV which is sensitive to the B-field in the region well
outside the torus of the Crab Nebula.

The constant B-field model offers the opportunity to cross calibrate
the IACT measurements with \emph{Fermi}/LAT observations. For the
IACTs an energy scaling factor $s_\mathrm{IACT}$ is introduced to
correct the measured energy $E_\mathrm{meas}$ to a common energy scale
$E = s_\mathrm{IACT}\times E_\mathrm{meas}$. This procedure reduces
the uncertainty of the energy scale to that of the \emph{Fermi}/LAT
\cite{Meyer2010}. The scaling factor is determined via a
$\chi^2$-minimization and is found to be $s_\mathrm{IACT} = 1.050 \pm
0.003$.

\clearpage

\end{document}